# Pink Work

## Same-Sex Marriage, Employment and Discrimination[♠]

Dario Sansone[1]

This version: 17 July 2018

**JOB MARKET PAPER**

The most recent version of the paper and the Online Appendix are available on my [website](#).

**Abstract**

This paper analyzes how the legalization of same-sex marriage in the U.S. affected gay and lesbian couples in the labor market. Results from a difference-in-difference model show that both partners in same-sex couples were more likely to be employed, to have a full-time contract, and to work longer hours in states that legalized same-sex marriage. In line with a theoretical search model of discrimination, suggestive empirical evidence supports the hypothesis that marriage equality led to an improvement in employment outcomes among gays and lesbians and lower occupational segregation thanks to a decrease in discrimination towards sexual minorities.

[♠] I am grateful to Garance Genicot, Francis Vella, Laurent Bouton, Jim Albrecht, Sarah Baird, Matteo Bobba, Stéphane Bonhomme, Mary Ann Bronson, Christopher Carpenter, Katherine Casey, Eric Edmonds, Luca Flabbi, James Foster, Adam Isen, Melissa Kearney, Thomas Lemieux, Rebecca Lessem, Arik Levinson, Lee Lockwood, Magne Mogstad, Hiren Nisar, Mariacristina Rossi, Andrew Shephard, Ben Solow, Allison Stashko, Eric Verhoogen, Alwyn Young, participants to the SAGE conference, IZA Summer School and seminar participants at American University and Collegio Carlo Alberto for their helpful comments. The usual caveats apply.

[1] Georgetown University (Department of Economics, ICC 580, 37th and O Streets, NW, Washington DC 20057-1036). E-mail: ds1289@georgetown.edu

# 1. Introduction

In 1895, Oscar Wilde was incarcerated because of "the love that dare not speak its name". Alan Turing, the father of computer science and artificial intelligence, was prosecuted for homosexual acts and forced to undergo chemical castration in 1952. In 1996, the U.S. federal government approved a law defining marriage as the union of one man and one woman. And yet, the last twenty years have seen exceptional improvements in LGBT rights: the Netherlands became the first country in the world to allow same-sex marriage (SSM) in 2001, homosexuality became legal in all U.S. states in 2003 (*Lawrence v. Texas*), while the first state in the U.S. to legalize SSM was Massachusetts in 2004. Finally, in 2015 the U.S. Supreme Court legalized same-sex marriage in all U.S. states (*Obergefell v. Hodges*).

Despite these advancements, little is known about the impact of these reforms on same-sex couples. At the same time, antigay sentiments might still be – although underreported - widespread (Coffman et al., 2017). Given the growing proportion of individuals identifying themselves as non-heterosexuals (7.5% according to Gates, 2017), it is important to understand how SSM legalization impacted the welfare of LGBT members and whether these legal changes shifted social norms and attitudes towards sexual minorities. In particular, given the discrimination from employers, consumers and coworkers commonly experienced by gays and lesbians, it is necessary to investigate whether SSM legalization led to improvements for sexual minority workers. Such an empirical evidence would provide guidance to policymakers as they continue to face conflicts over LGBT issues both in the U.S. (Economist, 2018) and Europe (Gillet, 2018).

Thus, this paper analyzes how the legalization of SSM affected gay and lesbian couples in the labor market. The empirical analysis in the first part of the paper exploits variations across U.S. states in the different timing of SSM reforms to show that this policy was associated with higher employment among same-sex couples: following SSM legalization, individuals in gay and lesbian couples were more likely to be employed and to work full-time. They also increased the total number of hours worked per week, while there was a reduction in the difference in hours worked between the household head and the partner. Male same-sex couples also experienced an increase in hourly earnings. These results are robust to several tests regarding parallel trends, timing of the reforms, migration, sample compositional changes and measurement errors.

The second part of the paper investigates potential mechanisms behind these results. After providing a theoretical framework using a search model with minority workers and prejudice employers, additional suggestive evidence from survey data and Google searches is provided to support the hypothesis that SSM legalization reduced discrimination based on sexual orientation, thus increasing the labor supply and demand of gay and lesbian workers. In line with this theory, there is also a decline in occupational segregation after the legalization of SSM. Alternative explanations, including changes in fertility, assortative matching and homeownership, are ruled out.

This paper contributes in several ways to the literature in household and labor economics. First and foremost, it exploits this unique opportunity provided by a large policy change in which the availability of marriage was extended to sexual minorities in order to investigate how access to such an institution affected several economic outcomes of interest. Historically, the only other analogous legal case is the invalidation of laws prohibiting interracial marriage after *Loving v. Virginia* in 1967 (Fryer, 2007), which has been shown to have led to an increase in mixed-race births (Fryer et al., 2012).

This analysis is also linked to previous studies on the impact of unilateral divorce laws on women. In particular, Stevenson and Wolfers (2006) argued that these reforms led to a reduction in female suicide and domestic violence, while Stevenson (2007) found that couples in states that allowed unilateral divorce were more likely to have both spouses employed in the labor force. Increases in female labor supply following changes in divorce laws have been documented both in the U.S. (Stevenson, 2007) and in Europe (Bargain et al., 2012). In contrast with these studies, this paper does not find evidence that SSM legalization led to a symmetric negative reduction in employment among gays and lesbians. This is rather surprising since, if anything, access to marriage led to changes in taxes, health insurances and adoptions that could have discouraged individuals in same-sex couple to be both employed.

This study is related by construction to the more general literature on sexual minorities. Despite the scarcity of data, economic analysis of LGBT issues started thanks to the pioneering works by Klawitter and Flatt (1998), as well as Badgett (2001). Later on, using mainly data from the U.S. Census, Black et al. (2007) provided a first glance at the economic lives of same-sex couples. The vast majority of the research on this topic has so far focused on the effect of sexual orientation on earnings using observational data (such as in Plug and Berkhout, 2004; Carpenter, 2007; (Clarke and Sevak, 2013), or correspondence experiments (see for instance Weichselbaumer, 2003; and Drydakis, 2009). As summarized by Klawitter (2015), most of these studies found that gay men earn less and lesbian women earn more than their heterosexual counterparts. More recently, Aksoy et al. (2018) documented the existence of a gay class ceiling, i.e. lower access to high-level managerial positions for sexual minorities.

Some researchers have also started to look at the effects of SSM legalization. Aldén et al. (2015) is the most related to the research question in this paper. The authors found suggestive evidence that entering into a registered partnership in Sweden was negatively associated with individual earnings for gay men, and positively related to fertility rates among lesbian women. In addition to this, Aksoy et al. (2018) argued that SSM legalization in Europe was followed by an improvement in attitudes towards sexual minorities, while Tankard and Paluck (2017) emphasized improvements in social norms following *Obergefell v. Hodges*. Reductions in sexually-transmitted diseases (syphilis) after SSM legalization have been found by Dee (2008) and Francis et al. (2012). This paper extends this literature by analyzing the impact of SSM legalization on several additional economic outcomes such as employment, earnings, occupational choices and self-employment.

Last but not least, the findings highlighted in the empirical analysis support the expansion of marriage equality. Homosexuality is illegal or barely tolerated in most countries outside Western Europe and North America. Previous research have pointed out that there is no effect of SSM legalization on heterosexual behavior (Badgett, 2009), including non-marital sex (Francis et al. 2012), marriage, divorce and extramarital birth rates (Dillender, 2014; Trandafir, 2015), thus dispelling any concerns about the potential destruction of traditional family values. This paper shows instead that a more tolerant environment can increase participation of gay and lesbian individuals in the labor market, thus offering an additional economic justification to the expansion of LGBT rights.

## 2. Institutional context

This Section provides a summary of the historical process that led to same-sex marriage legalization in the United States. A more detailed timeline and discussion of LGBT laws is available in the Online Appendix A.

The campaign for marriage equality in the United States started - with rather limited initial results - in the 1970s. In 1972, the U.S. Supreme Court denied appeal in *Baker v. Nelson*, a case in which the Minnesota Supreme Court ruled that the state statute limiting marriage to opposite-sex couples does not violate the U.S. Constitution. In 1973, Maryland became the first state to introduce a ban on same-sex marriage in its statute.

In 1993, the Supreme Court of Hawaii ruled that the prohibition on same-sex marriage violates the state constitution (*Baehr v. Miike*). The response at the state and federal level was immediate. In 1996, President Bill Clinton signed the Defense of Marriage Act (DOMA): a law defining marriage for federal purposes as the union of one man and one woman, and allowing states to refuse recognition of same-sex marriages granted in other states. Between 1994 and 1998, several states included a ban on same-sex marriage in their statute. For the first time in U.S. history, voters in Alaska and Hawaii approved constitutional bans to same-sex marriage in 1998.

Despite these backlashes, substantial progresses towards marriage equality were achieved between the end of the XX century and the beginning of the XXI century. In 1999, California introduced domestic partnerships, while Vermont became the first state to introduce civil unions in 2000. Massachusetts was instead the first U.S. state to legalize same-sex marriage in 2004, followed by Connecticut in 2008, and Iowa in 2009. Vermont also became the first state in 2009 to legalize same-sex marriage by statute instead of following a court decision.

All these changes generated reactions in other states: 26 states introduced bans to same-sex marriage in their constitution between 2004 and 2008. California swung several time over the years between legalizing and banning domestic partnerships, civil unions and same-sex marriage. Nevertheless, in 2013 the U.S. Supreme Court deemed Section 3 of DOMA unconstitutional (*United States v. Windsor*), thus forcing the U.S. government to recognize marriage unions performed in states that allowed such marriages, and to extend marriage-related federal benefits to same-sex married couples. Finally, same-sex marriage became legal in all the states after the U.S. Supreme Court decision in *Obergefell v. Hodges*.

## 3. Data

### 3.1 The American Community Survey

The main dataset used in the empirical analysis is the version of the American Community Survey (ACS) publicly available through IPUMS-USA (Ruggles et al., 2017). As described in Lofquist (2011), the ACS is a nationally representative survey containing demographic, economic, social, and housing data. It is conducted every year and has an average annual sample size of around 3 million addresses in the United States and Puerto Rico. The sample size has increased over time: the ACS 2000 had about a 1-in-750 national random sample of the population, which increased to around 1-in-240 from 2001, while from 2005 the ACS has surveyed 1% of the US population (IPUMS, 2017). The addresses include not only housing units, but also group quarters: e.g., nursing facilities and prisons.

The ACS is a mandatory survey. Although nobody has been prosecuted for not responding to the survey (Selby, 2014), this approach significantly increases self-response rate and the quality of the data: around 95% of the US counties are deemed to have acceptable quality data (U.S. Census, 2017).

The advantage of using this dataset is that, since the 1990, the U.S. Census Bureau has given respondents the option of classifying a member of the household as "unmarried partners" when asking about his or her relationship with the household head. In other words, roommates and unmarried partners have been treated as two separated categories starting from the 1990 Census. Therefore, it is possible to identify unmarried same-sex couples in the ACS. Furthermore, same-sex couples have been allowed to report their marital status since 2012.

Most of the empirical analysis focuses on household heads aged between 18 and 65 with married or unmarried partners. The household head is defined as the person who owns or rent the house, apartment or mobile home. If there is no such person, the first person listed can be any adult living in the household.

Even if the terminology is debatable, for simplicity the words "gay couple", "gays" and "male same-sex couple" are used as synonyms throughout the paper. The same logic applies to female same-sex couples and opposite-sex couples.

### 3.2 Descriptive statistics

In 2016, the last year available, same-sex couples represented 1.5% of all the unmarried and married couples in the sample. This is a substantial increase from 2008, the first year used in the empirical analysis, when same-sex couples represented only 0.9% of the sample. The percentage of female same-sex couples has always been slightly higher: in 2008, more than 51% of same-sex couples were composed by women, while such percentage in 2016 was around 50.5%. Among same-sex couples who decided to report their status in 2016, 52% of gays and 55% of lesbians were married (compared with 89% of opposite-sex couples).[2]

When looking at employment outcomes, it is interesting to note that in 67% of male same-sex couples both partners were working in the week preceding the interview, a higher percentage than among female same-sex couples (66%), unmarried opposite-sex couples (60%), and married opposite-sex couples (57%). These statistics are qualitatively similar to those reported in Black et al. (2007) using Census data. Even when focusing on young couples with children, same-sex couples were more likely to have both partners working (61% for gays, 64% for lesbians) than opposite-sex couples (60% for married, 53% for unmarried).

Similar differences are found among full-time workers: 49% of male same-sex couples had both partners usually working at least 40h/week in the 12 months preceding the interview, compared with 44% of female same-sex couples, 40% of unmarried opposite-sex couples, and 37% of married opposite-sex couples.

### 4. Econometric framework

The main econometric specification relies on the following difference-in-difference strategy:

$$y_{ist} = \beta SSM_{st} + \delta_s + \alpha_t + \tau_{ts} + \tau_{ts}^2 + x'_{st}\gamma_1 + x'_{ist}\gamma_2 + \varepsilon_{ist}$$

Where $y_{ist}$ is the relevant labor market outcome for individual (or household) *i* living in state *s* at time *t*. The coefficient of interest is $\beta$. $SSM_{st}$ is an indicator equal to one if the individual lived at time *t* in a state *s* where same-sex marriage had been legalized. The specification includes both state fixed effects ($\delta_s$) and year fixed effects ($\alpha_t$). As described in Angrist and Pischke (2014), the common trends assumption can be relaxed by controlling for state-specific linear and quadratic trends ($\tau_{ts} + \tau_{ts}^2$). It is also possible to include time-varying state-level controls $x'_{st}$, as well as individual-level controls $x'_{ist}$. State-level controls are important to take into account potential omitted-variable bias. While running a state-level regression with weights for population would give the same point estimates, the inclusion of individual-level controls is useful to increase precision (Angrist and Pischke, 2009). Since gay and lesbian couples may differ in their behavior, this specification is estimated for male and female same-sex couples separately when suspecting heterogeneities. Given the relatively large number of states (51, including the District of Columbia), it is possible to consistently estimate an arbitrary variance-covariance matrix: standard errors have been clustered at the state level to allow any correlation pattern within states over time (Bertrand et al., 2004).

The state-level controls $x'_{st}$ are similar to those included in Stevenson and Wolfers (2006): unemployment rate, income per capita, racial and age composition, proportion of state population with positive welfare (public assistance) income. In addition to these, $x'_{st}$ also includes the state heterosexual cohabitation rates, i.e. the proportion of different-sex couples who classify themselves as "unmarried partners", since this variable may capture higher levels of openness to SSM legalization (Badgett, 2009).

---

[2] Online Appendix B reports additional sample sizes and summary statistics for same-sex and opposite sex couples.

The individual and household controls $x'_{ist}$ include demographic characteristics of the household head and partner: age, education, race and ethnicity. Given its return to the labor market, this set of controls also includes whether the household head and the partner spoke English. One more reason because it may be important to control for age and education of the respondents is to account for match quality. Indeed, marriage laws may affect the quality of couple matches (Stevenson, 2007): being able to marry implies that it is more difficult to exit from a relationship, thus individuals may be reluctant to risk on a high-variance match. This would imply a higher match quality. On the other hand, marriage legalization would increase the benefits of entering into a relationship, which may lead individuals to become less selective. In other words, if there are advantages from getting married, individuals may lower their minimum threshold for match quality in order to start benefiting from marriage early on. In both cases, legalizing SSM may affect match quality, which may in turn affect labor market decisions.

Similarly, $x'_{ist}$ also includes the interaction between education of the household head and partner, as well as the age interaction, since positive assortative matching may affect specialization and labor market decisions through bargaining power and comparative advantages. For instance, if both partners are highly educated and have similar age, they are more likely to have equal weights in intra-household decisions.

## 5. Effect of SSM on employment

The estimates from the difference-in-difference model described above are presented in this section, while the main issues which may prevent a causal interpretation of these findings are discussed in Section 6. Evidence supporting the hypothesis that lower discrimination is the main mechanism driving the positive impact of SSM legalization on employment is presented in Section 7, and alternative potential channels are ruled out in Section 8. It is worth stressing that the difference-in-difference model estimates the impact of SSM legalization, not the impact of getting married. In other words, it is possible to estimate an intention-to-treat (ITT) effect, not an average treatment effect of marriage.

### 5.1 Main results

Table 1 shows the estimated effect of SSM legalization on the probability that both partners were working in the week preceding the interview. Columns 1 and 2 include only male same-sex couples, while Columns 3 and 4 focuses on female same-sex couples, and all these couples are combined in Column 5. Even after including state controls to account for time-varying factors potentially related with the legalization of SSM (Columns 2 and 4), the estimated coefficients are statistically significant and very similar for both gays and lesbians at around 2.3 percentage points. A similar estimate (2.4 percentage points) is obtained when looking at the effect on all same-sex couples (Column 5).

It is important to emphasize that the magnitude associate with SSM legalization is comparable to the impacts of other related policy reforms. In fact, Stevenson (2007) found an increase of 2 percentage points in the probability of both spouses being employed full time, and an increase of 2.4 percentage points in the probability that the wife was working following the introduction of unilateral divorce laws. Similarly, Bailey (2006) found an increase of 2-4 percentage points in the labor force participation rates of women aged 26 to 35 following the introduction of the pill. In contrast, several studies found very limited impact of childcare policies on labor supply (Fitzpatrick, 2010; Black et al., 2014).

### 5.2 Additional employment outcomes

As evident from Table 2, the effect of SSM legalization on employment was substantial also when looking at other related outcomes. This policy reform was associated with an increase of around 1.3 hours in the amount of time spent working weakly by the household head and the partner (Column 1), as well as an increase in the probability of both the household head and the partner working at least 40 hours per week (Column 2), or at least 30 hours per week (Column 3). In addition to this, the difference in the number of hours worked weekly by both partners got smaller (Column 4).

These findings are reassuring since these questions were asked considering the twelve months preceding the interview, while the employment status information examined in Table 1 was in regard to the week before the survey was conducted. Therefore, the estimated positive impact of SSM on labor market outcomes is found both at the intensive and extensive margin and it is not sensible to the time frame used to elicit employment. Furthermore, also in these cases the magnitudes are similar to those of other policy reforms: for instance, Bailey (2006) found an employment increase of 1.5-2.3 hour/week among women aged 26 to 35 after the introduction of the pill.

### 5.3 Earnings

Marriage is usually associated with an earning premium for heterosexual male workers. This may be due to increase productivity following intra-household specialization or behavioral changes (e.g., higher reliability and loyalty), employer discrimination in favor of married men, or positive selection into marriage. The magnitude and the mechanisms behind such marriage premium have been extensively debated in the literature (see for instance Ginther and Zavodny, 2001; Antonovics and Town, 2004; Dougherty, 2006).

On average, male employees in same-sex couples earned more than male workers in opposite-sex couples during the time period considered ($22 per hour vs $19.2). This is mainly due to the fact that those respondents were more educated: 57% of male respondents in same-sex couples had a Bachelor's degree or more, compared to 38.3% of men in opposite-sex couples. There was also a small gap between women in same-sex and opposite-sex couples ($18.5 vs. $17.9). In all types of couples, married employees earned more than unmarried individuals. The raw marriage premium was 15% for men in same-sex couples, 21% for women in same-sex couples, 40% for men in opposite-sex couples, 42% for women in opposite-sex couples.[3]

A similar increase in earnings for gays and lesbians may incentivize them to work more, thus supporting the results regarding employment in Tables 1 and 2. Using the same identification strategy implemented in the previous sections, Table 3 reports the difference-in-difference estimates of the effect of SSM legalization on earnings among male same-sex couples (Column 1), female same-sex couples (Column 2), and all same-sex couples jointly (Column 3). The coefficients are positive in all specifications, but only significant when looking at gay men. The coefficient for women is not precisely estimated, while the one for men implies an increase of 2.8% in the earnings of gay employees compared to gay workers living in states that had not legalized SSM yet. One possible explanation for this different effect among gay and lesbian couples might be that employers expect married individuals to be monogamous, thus reducing the risk of contracting HIV – more widespread among male homosexuals – and leading to better health, lower health insurance premiums, and higher wages.[4]

### 5.4 Extensions

The Online Appendix D includes several extensions. The conclusions from Table 1 and 2 are still valid after extending the time period considered (Section D.1). Qualitatively similar results are also obtained when examining the employment outcomes of individuals rather than couples (Section D.2). Furthermore, results from Table 1 are robust to restricting the sample to prime age workers (Section D.3.1).

Additional heterogeneities are investigated in Section D.3.2. There is some evidence that the gains from SSM legalization were higher among more marginalized groups. Indeed, the positive impact of this policy reform on employment was higher among lesbian couples without a Bachelor's degree or with medium-low household income. The probability of being employed after SSM legalization increased more for lesbian couples in Republican states than in Democratic ones. Black gay couples also seem to have particularly benefited from SSM legalization. This result is consistent with the idea of intersectionality and multiple discrimination (see for instance Brewer et al. 2002): being part of more than one minority group may have had further negative effects in the labor market, so a lower discrimination towards sexual minorities following SSM legalization could have benefited racial minorities the most given the intersection between race and sexual orientation.

---

[3] See also Section D.14 in the Online Appendix. The marriage premium is still detectable when comparing married and unmarried individuals in their prime age (25-55) or with children. This is true for male same-sex couples, female same-sex couples, and opposite-sex couples.

[4] Similar results are obtained once wages are adjusted for inflation, if part-time workers are excluded from the sample (done to disentangle the marriage premium from the part-time penalty), or using a triple-difference model (Section D.14 in the Online Appendix). It is also worth stressing that the interpretation of marriage premium used here is rather loose. The estimated coefficients measure the impact on earnings of SSM legalization, not of actually getting married. So they are intent-to-treat (ITT) estimates. It is not possible to obtain LATE estimates through IV since SSM legalization is not a valid instrument: it can affect earning also through different channels other than marriage.

The estimated impact of SSM legalization on employment is instead qualitatively similar between couples in urban and rural areas (Section D.3.3), while Section D.3.4 shows that the estimates in Table 2 are qualitatively similar when examining gays and lesbian couples separately, although not always significant when focusing on male same-sex couples only. In addition to this, since the total number of hours worked by the household head and the partner examined in Table 2 (Column 1) is a continuous variable, it is possible to analyze the impact of SSM legalization over the distribution of this outcome variable. As reported in Section D.3.5, the estimated effect of SSM is higher at the 0.25 quartile than at the 0.75 quartile. These results are in line with the lower gap between partners shown in Table 2 (Column 4): it suggests that unemployed and part-time workers experienced the largest increased in hours worked, thus reducing within-couple differences. Finally, Section D.4 shows that SSM led not only to higher employment, but also to higher labor force participation.

## 6. Threats to the identification strategy

As usual with the difference-in-difference model, there are several potential issues which may undermine the identification strategy. This section aims at discussing and ruling out the main threats. All related tables are included in the Online Appendix D.

### 6.1 Data Limitation

One key issue when dealing with same-sex couples is measurement error: indeed, a low rate of random error in a large group (opposite-sex couples) may create issues in the estimates of a small group (same-sex couples). As a result, there is the risk that several same-sex couples may actually be misidentified opposite-sex couples.

Section D.1 in the Online Appendix discusses at length this topic. To minimize such errors, observations with imputed sex and relation to the household head have been dropped (as suggested in Black et al., 2007). Moreover, the US Census Bureau implemented several changes between 2007 and 2008 to address this issue. The drop in the reported number of same-sex couples between these two years reported in Table B1 in the Online Appendix can be attributed to these changes, which have resulted in more reliable estimates of same-sex couple households (U.S. Census, 2013). Therefore, only observations from 2008 have been considered in the main empirical analysis.

Older respondents in opposite-sex couples are more likely to be misclassified as same-sex couples due to their lower levels of familiarity with issues and terminology pertaining to same-sex couples (Lewis et al., 2015). As already discussed, the main results are robust to excluding these older workers (Appendix D.3.1), thus confirms that these findings are not driven by measurement errors. Similarly, since the number of married opposite-sex couples is much larger than that of unmarried opposite-sex couples, married same-sex couples are much more likely to be misclassified opposite sex couples than unmarried same-sex couples (O'Connell and Feliz, 2011). Nevertheless, excluding these married couples – identified in the ACS since 2012 - from the analysis provides very similar estimates from those in Tables 1 and 2.[5]

The source of measurement error is well-known in this case: opposite-sex couples misclassified as same-sex couples. Therefore, it is possible to test how sensible are the estimates in Table 1 to such misclassification by randomly adding some opposite-sex couples to the sample of same-sex couples. As expected given the null impact of SSM legalization on opposite-sex couples (see Section 6.6 below), the estimated coefficient of SSM gets smaller and smaller as the misclassification error increases. Nevertheless, it remains positive and statistically significant up to plausible levels of misclassifications, thus implying that the main conclusions are not extremely sensible to this kind of measurement error.[6]

It is also important to stress that, despite these issues, the U.S. Census and the ACS remain some of the largest and most reliable data on same-sex couples. In fact, the across-metropolitan distribution of gay couples in the 1990 Census data line up extremely well—with a correlation of nearly 0.90—with AIDS deaths in 1990, a year during which AIDS deaths were predominately concentrate among gay men (Black et al., 2000). Using health data, Carpenter (2004) showed that same-sex unmarried partners were indeed behaviorally gay and lesbian, i.e. they exhibited sexual behaviors that were different from opposite-sex couples. Moreover, the author was able to replicate the findings on lower household income among lesbian couples and earning penalty experienced by gay workers highlighted in previous studies based on Census data, thus confirming that data limitations in the U.S. Census were not driving the results.

Another advantage of ACS is that a third of the households use Computer Assisted Telephone or Personal Interviews (CATI or CAPI). In such interviews, respondents are asked to verify the sex of their same-sex husband/wife, thus reducing such miscoding (Gates and Steinberger, 2007). There are other surveys with contains information on sexual orientation, e.g. the General Social Survey (GSS). However, these alternative data sources have limited sample sizes, thus the ACS is the only survey which allows to exploit the year-to-year changes in SSM laws between states. On the other hand, the main disadvantage is that it is possible to identify in the ACS only gay and lesbian individuals in same-sex couples, not singles. Furthermore, there is no information on sexual behavior, so it is not possible to detect members of opposite-sex couples who are bisexuals.

### 6.2 Triple difference and other contemporaneous reforms

It may be argued that same-sex couples living in Massachusetts are not comparable to those living in more conservative states such as Mississippi. First, it is important to remember that the difference-in-difference model requires only parallel trends, not similar baseline characteristics. Moreover, the specifications presented in the previous section also include several state and individual controls. As an additional robustness check, it is possible to estimate a triple-difference model, i.e. comparing same-sex and opposite-sex couples within the same state over time. The estimated impact of SSM legalization is still positive, statistically significant, and with magnitude close to the coefficients in Table 1.[7] Since this estimate is obtained by comparing same-sex and opposite-sex couples *within* the same state, it also suggests that the positive impact found in the difference-in-difference estimates when comparing same-sex couples *between* states was not due to backlashes and negative employment outcomes in states that had not legalize SSM, but rather to actual improvements in states that legalized SSM.

---

[5] See Section D.1 in the Online Appendix. Section 7 below further discusses the implications and interpretation of a significant impact of SSM legalization on unmarried couples.
[6] Estimates reported in Section D.1 in the Online Appendix.

In addition to this, it is reassuring to note that the results in Table 1 are not driven by other contemporaneous reforms or by changes in only one state. As shown in Table 5, the estimate impact of SSM is robust to the inclusion of other policy indicators as controls, such as whether the state introduced anti-discrimination laws in the same period. Moreover, the contemporaneous introduction of other reforms would have been a major concern if the identification strategy had relied on only two states. For instance, just comparing Massachusetts to a similar state before and after 2004 would have been problematic since Massachusetts implemented a health reform around the same period. Nevertheless, this is less of an issue when using all U.S. states since each state has a lower weight than in a pair comparison. In addition to this, Section D.6 in the Online Appendix shows that the estimates in Table 1 remain positive and statistically significant even when excluding one state at a time.[8] This is particularly reassuring since several findings published in top economic journals have been found to be extremely sensible to outliers (Young, 2017).

## 6.3 Migration and compositional changes

Between-state migration may have also changed the geographical composition and distribution of same-sex couples, thus leading the difference-in-difference model to compare different populations over time. Nevertheless, there is no evidence that same-sex couples massively moved to states that legalized SSM.[9] In line with these results, Stephens-Davidowitz (2013) found low migration rates from less tolerant states using data from Facebook.

The estimation and interpretation of the impact of SSM legalization may also depend on how same-sex couples have been identified: individuals may differ over time and between states in their propensity to be in a homosexual relationship and to be open about it. Indeed, despite the anonymity guaranteed by the U.S. Census, some individuals may have decided not to truthfully report their sexual orientation.[10] In order to include also these "closeted" couples, Table 4 has a similar structure to Table 1, but estimates the impact of SSM legalization on the probability that both partners are working among same-sex married couples, unmarried couples, and households with two same-sex roommates.

---

[7] Section D.5 in the Online Appendix reports and further discusses the estimates from the triple-difference model.
[8] This test is similar to the one implemented in Angelini et al. (2016) when analyzing the impact of divorce laws.
[9] Section D.7 in the Online Appendix reports the estimated coefficients.
[10] This behavior is similar to the historical manipulations of racial appearance and the attempts to "pass" as white among Americans with African ancestry (Nix and Qian, 2015).

Only couples in which one household member was listed as roommate and had the same sex of the household head have been included among same-sex roommates since in case of multiple roommates it was not possible to infer the identity of the household head's partner (if any). Roommates were not considered when the household had also a spouse or unmarried partner. Moreover, only couples whose household head was aged between 30 and 60 have been considered to reduce the risk of including cohabitating students or older individuals living with non-relatives.

It is quite interesting to note that there are large differences between states when looking at the proportion of opposite-sex and same-sex couples (married, unmarried or roommates). For instance, 98.9% of these couples are opposite-sex in Mississippi. In contrast, opposite-sex couples represent 89% of all the couples in DC. The proportion of same-sex roommates is similar to that of same-sex married/unmarried couples in less tolerant states such as Alabama, Mississippi, Texas and Louisiana, while it is smaller in more LGBT-friendly states such as Massachusetts, Vermont, New York and DC.[11] This result provides evidence supporting the hypothesis that individuals in same-sex relationships are more likely to report being roommates when they prefer not to disclose their sexual orientation.

The coefficient of SSM legalization remains positive, statistically significant and with a magnitude equal to the one estimated in Table 1 Column 5 even when including same-sex roommates (Table 3 Column 4). Therefore, the main results are not driven by SSM legalization affecting the composition of the sample, i.e. by changing how many same-sex couples decided to be open about their sexuality rather than classifying themselves as roommates.

In line with this result, there is no evidence that SSM legalization led to a substantial change in the demographic composition of same-sex couples within each state. Section D.9.1 in the Online Appendix reports the estimates from the difference-in-difference specifications using household head's demographic characteristics as dependent variables. SSM legalization did not substantially affected the ethnic, age and language composition of same-sex couples.

However, it seems that more individuals without tertiary education decided to be open about their sexuality. If anything, since education is positively related with employment, this change should lead to a downward bias in the estimates in Tables 1 and 2. Indeed, 87% of the household heads in same-sex couples with a Bachelor's degree or higher were employed at the time of the survey, while the same figure is 76% among those without tertiary education. Given this sample variability due to education, one way to construct some bounds for the impact of SSM legalization is to restrict the sample only to individuals with a certain educational level, thus limiting compositional changes by construction. As shown in Section D.9.2 in the Online Appendix, couples in which both partners had at least a Bachelor's degree were 2.2 percentage points more likely to be employed following SSM legalization. The same coefficient goes up to 2.4 percentage points when examining couples in which only one partner had achieved a higher educational level, while the upper bound is 3.1 percentage points among couples without tertiary education.

---

[11] See also Section D.8 in the Online Appendix

Similarly, as reported in Appendix D.11, results in Table 1 are robust to the exclusion of individual controls ($x'_{ist}$). The coefficient of SSM legalization remains positive and statistically significant when examining the probability of both partners working. This finding is thus in line with the assumption that the main conclusions are not driven by compositional changes.

To further reassure about the validity of the estimates, it is possible to combine difference-in-difference with matching. This extension compares individuals in treated stated following SSM legalization with comparable individuals in treated states at the baselines, as well as with comparable individuals in control states at the baseline and after the treatment. Merging these two methods ensures that similar individuals are compared across time and space, thus verifying that sample compositional changes are not pivotal. As reported in Section D.9.3 in the Online Appendix, the estimated impact of SSM legalization when augmenting the difference-in-difference model with kernel weights computed from propensity scores remains positive, statistically significant and with magnitude close to the coefficients shown in Table 1.

### 6.4 Anticipation and parallel trends

Section D.10 in the Online Appendix discusses why several past turnarounds and abrupt law repeals make it is unlikely that same-sex couples started changing their labor market decisions before the actual legalization of SSM. Moreover, the estimated coefficient of an additional lead indicator ($SSM_{st+1}$) is statistically insignificant and close to zero in magnitude. This result rules out both any anticipation effect and the hypothesis that improvements in the labor market were actually driven by changes in attitudes among the general population *before* SSM legalization. In other words, the statistically insignificant coefficient of $SSM_{st+1}$ does not support the idea that - after controlling for linear and quadratic trends - attitudes towards sexual minorities among heterosexuals improved before the legalization of SSM, and that they affected both the change in policy and the variation in employment.

More generally, adding up to three lead operators to the specifications in Tables 1 and 2 still results in statistically insignificant coefficients, while the contemporaneous effect of SSM legalization remains significant.[12] These findings not only confirm that no changes occurred before the policy reform (after conditioning on linear and quadratic state-specific trends), but also support the parallel trend assumption in the difference-in-difference model. Indeed, since all these additional lead variables are indistinguishable from zero, it is possible to assume that trends in employment among same-sex couples were similar among U.S. states before the legalization of SSM.

---

[12] Section D.11 in the Online Appendix reports the estimated coefficients. It also shows that the coefficients of $SSM_{st}$ remain statistically significant after the introduction of additional leads and lags in the model. Moreover, it provides suggestive evidence that the impact of the policy was concentrated in the year SSM was legalized, with little effect before or after.

Finally, as reported in Appendix D.11, results in Table 1 are robust to the exclusion of the linear and quadratic state-specific trends ($\tau_{ts}$ and $\tau_{ts}^2$). The impact of SSM legalization on the probability of both partners working remain positive and statistically significant. This result thus provides additional evidence supporting the parallel trend assumption.

**6.5 Timing of the reform**

Another key concern is that the timing of this policy reform should not reflect pre-existing difference in state-level characteristics. First and foremost, unlike other policy reforms such as unilateral divorce laws (Stevenson, 2007), SSM legalization was primarily driven by state and federal courts' decisions. Therefore, the time of the legalization was less likely to depend on other time-varying state characteristics since judges were – arguably - less influenced by public opinion than policymakers. Indeed, state courts started to legalized SSM in Massachusetts, Iowa and Connecticut before 2010 even if pools did not show national popular support for SSM until 2011-2013 (McCarthy, 2017; Pew, 2017). For instance, when Massachusetts legalized SSM in 2004, only 36% of residents in New England did not oppose sexual relationships between two adults of the same sex.[13]

In addition to this, the main difference-in-difference specifications include several state-level variables which may have affected the legalization of SSM. As already mentioned, this set of controls includes the share of opposite-sex unmarried couples in the state. According to Badgett (2009), higher cohabitation rates may signal a higher level of openness towards sexual minorities and different family structures in the society, so taking this factor into account reduces concerns about the endogeneity of the policy reforms.

It is also worth pointing out that state fixed effects encompass all the time-invariant characteristics of the state, such as religion or political affiliation. This is the same argument used by Bailey (2006) when analyzing the impact of the pill on female labor supply to control for the fact that a strong Catholic lobby may have delayed the diffusion of birth control methods. The time span considered in the empirical analysis is rather short (2008-2016, 9 years), so it is likely that these variables did not change in this time period. Last but not least, all specifications include linear and quadratic state-specific time trends.

As an additional robustness check, Column 6 of Table 1 shows that the effect of SSM legalization on the probability that both same-sex partners were working remains positive, statistically significant, and with an even larger magnitude when restricting the sample to the years between 2012 and 2016. In this time period, most of the legal changes were driven by decisions from the U.S. Supreme Court, thus they were not influenced by local factors. It is also possible to restrict the sample size even more and consider only the 2014-2016 time period. In this specification the coefficient of SSM legalization is identified only through the *Obergefell v. Hodges* decision by the federal Supreme Court, a sentence independent of state characteristics. Nevertheless, the estimate remains positive and statistically significant. Its magnitude also increases up to 6 percentage points (Column 7 Table 1).[14]

---

[13] Source: GSS data, author's own calculation.

**6.6 Placebo tests**

As discussed in Athey and Imbens (2017), placebo analyses are the most common form of supplementary analyses conducted to shed light on the credibility of the primary analysis. In these analyses, the true impact of the regressor of interest on the pseudo-outcome is known to be zero, so the goal here is to test whether the identification strategies implemented in the primary analysis produce estimates close to zero when applied to such pseudo-outcome.

The estimated coefficients of SSM legalization are statistically insignificant and with tight confidence intervals around zero when focusing on opposite-sex couples.[15] Therefore, this null result from such a placebo test reinforces the validity of the main analysis. This finding is also in line with the work of Badgett (2009), Francis et al. (2012)), and Trandafir (2015) highlighting that SSM legalization did not affect heterosexual behavior. In other words, as also discussed in the Introduction, there is no evidence that SSM led to a deterioration of the American family.

**7. Discrimination**

Given the positive effect of SSM legalization on employment and earnings, it is worth investigating more in depth whether marriage equality actually led to a decrease in discrimination towards sexual minorities. Indeed, previous studies have emphasized the positive effect of lower discrimination on labor supply and employment among women and racial minorities (see for instance Leonard, 1990; and Collins, 2001), so it is possible that SSM legalization triggered the same mechanism.

This section describes the conceptual framework linking SSM legalization with discrimination and employment. It then collects a set of supplementary analyses to support this theory. The Online Appendix provides additional evidence using data on hate crimes (Section E.4) and on attitudes towards homosexuals (Section E.5). The findings in Tankard and Paluck (2017) are also in line with the mechanism highlighted in this section: the authors randomly assigned participants from Amazon Mechanical Turk to read either a positive or negative analysis of the U.S. Supreme Court decision on SSM before the actual ruling. Individuals in the treatment were more likely to report higher perceived support for gay marriage among Americans. Improvements in perceived social norms were also found in a longitudinal study when comparing participants' answers before and after *Obergefell v. Hodges*.

---

[14] In addition to this, Section D.12 in the Online Appendix also finds no evidence of a substantial omitted variable bias following (Oster (2017)). It also rejects the hypothesis that the effect of SSM legalization was different in states that legalized SSM following a state or federal court decision than in states that directly introduced the law in the state statute.

[15] See also Section D.13 in the Online Appendix.

## 7.1 Conceptual framework

There are different reasons which may explain discrimination towards sexual minorities. The classical theory is the one of taste-based discrimination developed by Becker (1957): some employers may dislike minority workers, thus making it more expensive to hire this kind of workers compare to equally productive heterosexual individuals. Furthermore, even in absence of personal prejudices, a profit-maximizer employer would discriminate if she expected that consumers prefer to interact with heterosexual workers, or that other employees prefer to work with heterosexual co-workers, and thus hiring a minority worker may reduce overall firm productivity and increase staff turnover.

The second leading theory is the one of statistical discrimination described in Arrow (1973): given the uncertainty about the actual productivity of a job candidate, employers may try to predict the quality of the worker from the (perceived) average productivity of minority workers. From this perspective, gay men may be discriminated because deemed less masculine as a group, a trait usually remunerated in several occupations. Moreover, employers could believe that gay men are more likely to have HIV, and that their health condition may affect their productivity (even if this is no longer true thanks to the new generation of anti-retroviral drugs) or that they might – though some mysterious ways – infect other workers. Lesbian women may be actually positively discriminated due to their perceived lower fertility, higher labor force attachment, and stronger personality (Patacchini et al., 2015), although the empirical evidence is far from clear-cut (Weichselbaumer, 2003). Second-order statistical discrimination stems instead from the (perceived) higher variance in productivity among minority workers (Klumpp and Su, 2013). Employers are less familiar with minority workers, so even if they perceived these workers to be on average as productive as heterosexual workers, they might be reluctant to hire them due to the higher uncertainty.

More recently, Pęski and Szentes (2013) based their model of discrimination on social norms rather than preferences or productivity reasons: heterosexual employers may discriminate minority workers because such behavior is tolerated, expected and deviations are punished by other heterosexual individuals.

SSM legalization may affect all these kinds of discrimination. First, this policy may shape preferences and change attitudes among employers, workers and consumers towards sexual minorities (Aksoy et al., 2018). These individuals may also realize that most homosexuals do not follow their idea of "gay sinful lifestyle", but rather they have the same aspirations to get married and have a family. Second, higher visibility of LGBT individuals may dissipate old cliché about gay men as all feminine and thus reduce statistical discrimination. As more homosexual workers are hired or decide to come out, employer would adjust their expectations about average productivity and variance for this group of employees. Given the time required to update employers' expectations, short-term decreases in discrimination would be mainly due to a decline in taste-based discrimination, while lower statistical discrimination should drive long-term trends.

Third, social norms might be affected and employers may realize that past discriminatory behaviors are no longer considered acceptable, and that having a diverse workforce is not punished, but actually valued.

Finally, the documented increase in employment among same-sex couples may be due not only to an increase in labor demand, but also to a higher labor supply. Indeed, through a feedback mechanism, more gay and lesbian individuals could decide to participate in the labor market since they envisage lower discrimination and thus their expected wages exceed their reservation wages. This mechanism is in line with the effect hypothesized by Neumark and McLennan (1995) when examining female employment: women may invest less in the labor market when expecting discrimination.

## 7.2 Theoretical framework

In order to formalize and clarify how SSM might have affected labor market outcomes for same-sex couples through a reduction in discrimination, this section presents a search model with minority workers and prejudiced employers by adapting and extending the work in Flabbi (2010a) and Flabbi (2010b). All the details and proofs are discussed in Section E.1 in the Online Appendix.

This model is a random search model set in continuous time with job destruction and no on-the-job search. There are two types of employers: prejudiced firms ($P$) and unprejudiced ones ($N$). The share of prejudiced employers is $p$. There are two types of workers: minority employees ($G$ for gay) and non-minority ones ($S$ for straight). The disutility incurred by prejudiced employers when hiring a minority worker is $d$. Workers can be in one of three different states: employment ($e$), unemployment ($u$) and non-participation in the labor market ($1 - l$).

The flow value of non-participation in the labor force is $z \sim Q(z)$, while $b$ is the value of unemployment. Unemployed workers and firms randomly meet following a Poisson process characterized by parameters $\lambda_G$ and $\lambda_S$ for minority and non-minority workers respectively. Once employer and worker meet, the match-specific productivity value $x \sim G(x)$ is revealed. If a match is realized, the wage $w(x)$ is determined through Nash-bargaining with the worker's weight equal to $\alpha \in [0,1]$. Finally, $\eta$ is the job-destruction rate and $\rho$ is the intertemporal discount rate.

Given these parameters, it is possible to write down the present-value of non-participating in the labor market $NP_J(z)$, of employment $V_J[w_{JI}(x)]$, and of unemployment $\rho U_J$:

$$NP_J(z) = \frac{z}{\rho}$$

$$V_J[w_{JI}(x)] = \frac{w_{JI}(x) + \eta U_J}{\rho + \eta}$$

$$\rho U_J = b + \lambda_J \left\{ p \int max[V_J[w_{JP}(x)] - U_J, 0] \, dG(x) + (1-p) \int max[V_J[w_{JN}(x)] - U_J, 0] \, dG(x) \right\}$$

For $J = G, S$ and $I = N, P$.

Wages are instead determined through Nash bargaining:

$$w_{JI}(x) = \alpha(x - dI_{\{G,P\}}) + (1-\alpha)\rho U_J$$

For $J = G, S$ and $I = N, P$.

Reservation values can be easily derived from these equations, and it is then possible to define the equilibrium.

*Proposition 1*: Given the exogenous parameters $\{\lambda_G, \lambda_S, \eta, \rho, b, \alpha, d, p\}$ and the distribution functions $G(x)$ and $Q(z)$, the unique steady state equilibrium is defined by the following three conditions:

$$\rho U_J = b + \frac{\lambda_J \alpha}{\rho + \eta}\left\{p\int_{\rho U_J + dI_{\{G,P\}}}^{+\infty}[x - dI_{\{G,P\}} - \rho U_J]dG(x) + (1-p)\int_{\rho U_J}^{+\infty}[x - \rho U_J]dG(x)\right\}$$

$$u_J = \frac{\eta}{\eta + \lambda_J\{p[1 - G(\rho U_J + d1_{\{G,P\}})] + (1-p)[1 - G(\rho U_J)]\}}$$

$$l_J = Q(\rho U_J)$$

SSM legalization may lead to a lower $d$ since employers would expect lower discrimination from coworkers and consumers. Moreover, employers could be less afraid of violating outdated social norms when acting within a supporting legislative framework. In addition to this, SSM might increase expectations of monogamy among same-sex male couples, thus reducing the risk of HIV, and leading to lower health insurance costs for employers. The comparative statics for such a change in $d$ are summarized in the following proposition.

*Proposition 2*: For any equilibrium previously defined, as the disutility from hiring minority workers $(d)$ decreases, the wage received by minority workers $(w_G)$ increases, their unemployment rate $(u_G)$ decreases, and their labor force participation rate $(l_G)$ increases.

In addition to these effects through direct changes in labor demand, direct changes in labor supply may occur since same-sex couples may expect lower discrimination following SSM legalization. This would result in larger efforts in the job search - a "feedback" effect similar to the one described in Neumark and McLennan (1995) - thus increasing the probability of meeting a firm.

*Proposition 3*: For any equilibrium previously defined, as the job arrival rate for minority workers $(\lambda_G)$ increases, their wage $(w_G)$ increases, their unemployment rate $(u_G)$ decreases, and their labor force participation rate $(l_G)$ increases.

These comparative statics are in line with the main results discussed in Sections 5 and 6, while the next sections provide evidence to support the hypothesis that SSM legalization reduced discrimination.

Finally, the model provides interesting results also regarding occupational segregation. Indeed, while non-minority workers are indifferent between prejudiced and unprejudiced employers since they are not treated differently, minority workers would prefer ex-ante to work for an unprejudiced firm, but they might end up working for a prejudiced employer if the match has a high enough productivity value. Consequently, the proportion of minority workers working for unprejudiced employers in equilibrium is:

$$P_{GN} = \frac{(1-p)[1 - G(\rho U_G)]}{p[1 - G(\rho U_G + d)] + (1-p)[1 - G(\rho U_G)]}$$

It is then possible to prove that, under certain functional form assumptions:

*Proposition 4*: For any equilibrium previously defined, as the disutility from hiring minority workers ($d$) decreases, occupational segregation ($P_{GN}$) declines.

### 7.3 Additional policy reforms

The first suggestive evidence supporting the hypothesis of lower discrimination following SSM legalization is provided by Table 5. This table extends the difference-in-difference model with male and female same-sex couples estimated in Table 1 (Column 5) by including additional policy indicators, that is, whether states introduced other policies affecting homosexual individuals in the time period considered.

In particular, these policies indicate whether and in which year a state introduced a constitutional ban on same-sex marriage (Column 2), legalized domestic partnership and civil union (Column 3), introduced anti-discrimination laws (Column 4), and allowed or prohibited adoptions by same-sex couples (Column 5). It is worth emphasizing that these results are only presented as suggestive evidence to reinforce the findings on SSM legalization. Indeed, while 48 states legalized SSM between 2008 and 2016, only between 1 and 7 states implemented one or more of these additional reforms.[16]

First, it is reassuring that the coefficient of SSM legalization remains positive and statistically significant in all specifications, thus minimizing any concern about endogeneity. Second, the impact of these other policy reforms is also consistent with the idea that employment increased because of higher tolerance signaled and caused by these laws. Couples living in states that introduced a constitutional ban on SSM or adoptions were less likely to be both working. On the other hand, there is a positive association between the legalization of domestic partnerships (or civil unions) and employment. Similarly, the coefficients of anti-discrimination laws and second-parent adoption are also positive.[17]

---

[16] See also Sections A.1-3 in the Online Appendix for the complete timeline of these reforms.

[17] As discussed in Section E.2 in the Online Appendix, there is no evidence of heterogeneity in the effect of SSM marriage between more or less tolerant states. Indeed, the interaction terms between SSM legalization and whether the state passed sexual

**7.4 Heterogeneities by relation status.**

If indeed SSM legalization led to lower discrimination based on sexual orientation, then all same-sex couples should have benefited, even those unmarried. This hypothesis is investigated in Table 6. Since the U.S. Census started allowing same-sex couples to classified themselves as married only from 2012, these specifications have been estimated using the ACS 2012-2016 waves. The usual caveat in this case is that marital status is endogenous, so results in this section are presented as suggested evidence, not causal inference.[18]

In line with the idea of a widespread impact among gays and lesbians, SSM legalization led to an increase in the probability that both partners were working also among unmarried same-sex couples. In fact, the coefficient associated with the policy indicator is positive in all specification, and statistically significant when focusing on lesbians (Column 2), or all same-sex couples jointly (Column 3). Consistently with the descriptive statistics presented in Section 3.2, the coefficient of marital status is negative and significant (probably due to different fertility rates between married and unmarried couples). However, married individuals benefited even more from the policy change: the interaction term between marital status and SSM legalization is positive and significant. Overall, the positive effect of SSM legalization compensate the negative effect of marital status.[19]

An alternative way to test the hypothesis that SSM legalization affected all same-sex couples, not only married ones, is to extend the sample to include also same-sex roommates. The estimated coefficients in Table 4 support the hypothesis that the benefits stemming from SSM legalization were not concentrated only among gays and lesbians who decided to get married. The interaction term between the policy variable and the roommate indicator is statistically insignificant for gays (Column 1), lesbians (Column 2), and when examining all same-sex couples jointly (Column 3). Thus, this suggests that the impact of SSM legalization was similar among same-sex couples and (potentially closeted) same-sex roommates.

**7.5 Evidence from Google Trends**

Google searches provide an interesting alternative data source to investigate changes in animosity towards gays. Indeed, Google data are a good proxy for socially sensitive attitudes since users are online, alone, and have an incentive not to lie in order to obtain what they are looking for: all these factors make it easier to express opinions on sensible topics such as race, health, or sexual practices

---

orientation anti-discrimination laws to protect private and/or public employees are not statistically significant. An alternative way to assess tolerance can be obtained by measuring the proportion of same-sex couples among all the married and unmarried couples in a given state and year. The coefficient of SSM legalization remains positive and statistically significant after including this additional state-level control, white its interaction term with the proportion of same-sex couples is not statistically significant.

[18] It is worth emphasizing that the interaction term between marital status and SSM legalization would still be consistently estimated if certain higher-order conditions were met (Bun and Harrison (2018)). (Nizalova and Murtazashvili (2016)) argued that the interaction term is consistently estimated also when the endogenous regressor (marital status) and the unobservables are jointly independent from the treatment variable (SSM legalization), but this assumption seems less realistic in this context.

[19] One possible explanation for these results is that employers discriminated less all homosexual employers, and that they had a preference for married individuals given a positive signal of reliability and "heteronormative" lifestyle.

(Stephens-Davidowitz, 2014). As discussed in Choi and Varian (2012) and Dergiades et al. (2015), it is possible to use Google Trends to compute a time series index of the volumes of queries entered by users into Google in each given U.S. state. This search intensity index is based on query shares normalized between from 0 to 100. A query share is the total query volume for a given search term(s) within a particular geographical region divided by the total number of queries in that region during the time period being considered. The maximum query share in the time period considered is set to 100. The following difference-in-difference model can therefore be estimated:

$$y_{st} = \beta SSM_{st} + \delta_s + \alpha_t + \tau_{ts} + \tau_{ts}^2 + x'_{st}\gamma + \varepsilon_{st}$$

Where $y_{st}$ is the search intensity for a given word in state *s* at time *t*. All the other regressors are defined as in the main sections. However, in this case data are available only at the state level, not the individual level, so the number of observations is substantially reduced and it is not possible to include individual controls as in the previous sections.

The first word used to approximate attitudes towards homosexuals is *Leviticus*. This is a book of the Bible and contains the reference "You shall not lie with a man as with a woman, this is an abomination" which has been often used to justify homophobia among Christians and Jews. As reported in Table 7, SSM legalization led to a statistically significant reduction in search intensity for this term of 2.1 points out of 100 (Column 1). Despite the relatively small sample size, this coefficient remains large, negative and significant also when including the additional policy indicators reported in Table 5 (Column 2). Since this decline may simply reflect lower media coverage and interest after the passage of the law, Column 3 includes two lagged operators: the decline in Google searches is negative and significant even two years after the legalization of SSM, thus reflecting a long-lasting decline in this kind of derogatory searches. Similar conclusions are obtained when controlling for the overall search intensity on LGBT topics (Column 4), as well as when including state-specific linear time trends (Column 5). The estimates remain negative but become too imprecise only when adding state-specific quadratic time trends (Column 6), arguably because the number of regressors (179) is too close to the number of observations (663).

Section E.3.1 in the Online Appendix includes also two lead operators ($SSM_{st+1}$ and $SSM_{st+2}$) in order to emphasize that searches for the word *Leviticus* did not start to decline before the legalization of SSM. This is consistent with the hypothesis discussed in Section 6.4 that changes in attitudes did not predate legal changes. Similar (although not always statistically significant) results are presented in the Online Appendix from the analysis of additional Google trends for both words with a negative connotation (*Sodomy* in Section E.3.2, *Faggot* in Section E.3.3) and a positive one (*Gay pride* in Section E.3.4).

### 7.6 Occupational segregation

Plug et al. (2014)) documented that gay and lesbian workers sort into tolerant occupations: by comparing twins with different sexual orientations, the authors found that gays and lesbians are less likely to work in occupations with prejudiced workers. Similarly, Black et al. (2007) noticed that gay workers are in occupations with a higher proportion of women than straight male workers, while the opposite is true for lesbian workers. Similar summary statistics are obtained from the ACS: male workers in same-sex couples are in occupations with a higher share of women than men in opposite-sex couples. Female workers in same-sex couples are instead in occupations with a higher share of men than women in opposite-sex couples. In addition to this, there are no differences in the average shares of women within occupation when comparing married and unmarried couples by gender and sexual behavior.[20]

As formalized in Proposition 4 in the above search model, a lower level of discrimination following the legalization of SSM might induce a shift of gay and lesbian workers towards historically less tolerant occupations (or to disclose their sexual orientation if already employed in these sectors). One way to test whether SSM legalization reduced occupational segregation is by investigating whether individuals in same-sex couples were employed in occupations with a lower share of female workers after these policy reforms.

Table 8 reports these estimates from a difference-in-difference model as the one introduced in Section 4. The dependent variable in Column 1 is a binary indicator equals to one if the respondent (either the household head or the spouse/partner) was employed in an occupation with a majority of female workers.[21] The coefficient associated with SSM legalization is negative (1.4 percentage points) and statistically significant.

Similar estimates are obtained from alternative specifications. The coefficient of SSM legalization remains negative and significant when restricting the sample to household heads only (Column 2). Each respondent reported her last occupation, but the estimates are similar when examining only individuals who were employed at the time of the interview (Column 3). Qualitatively similar results are obtained when examining the share of female workers within occupation as dependent variable rather than a binary indicator (Column 4). [22]

Related to this, it is interesting to investigate whether SSM legalization affected another main occupational choice: paid work vs. self-employment. The last column of Table 8 shows a decline in self-employment among same-sex couples following the legalization of SSM. These couples are 1.9 percentage points less likely to have at least one member of the household working for her own enterprise. This result is in line with the hypothesis that gays and lesbians shifted out of self-employment given lower expected discrimination from employers (and co-workers).[23]

---

[20] See the descriptive statistics in Section E.6.1 in the Online Appendix.

[21] Section C.1 in the Online Appendix clearly defines each dependent variable examined in this section.

[22] Section E.6.2 in the Online Appendix includes additional robustness checks: computing the share of women within occupation using weighted averages rather than unweighted ones leads to very similar estimates. Moreover, the decline seems to have been larger among lesbian workers than gay ones. Section E.6.3 presents instead additional empirical analyses using occupational data from GSS.

[23] Section E.6.2 in the Online Appendix includes additional robustness checks: qualitatively similar estimates are obtained when analyzing only whether the household head is self-employed, or when using an alternative definition of self-employment. Furthermore, the effect seems to be larger among male same-sex couples than female same-sex couples.

## 8. Alternative explanations

There are different alternative explanations for the relation between SSM legalization and the employment outcomes of gays and lesbians highlighted in the main analysis. This section starts with a major channel interrelated with labor market decisions – fertility – and then briefly discuss additional mechanisms.

### 8.1 Fertility

There is one pivotal reason which may have explained a *negative* impact of SSM legalization on employment: marriage may have incentivized intra-household specialization. Indeed, Becker (1973) identified the main advantages from marriage in the intra-household specialization: one person in the couple try to increase his or her productivity and earnings in the labor market while the partner specializes in the production of household commodities, that is, in goods that are non-marketable or transferable among different households, although they may be transferred among members of the same household, such as children. However, as stressed in Stevenson and Wolfers (2007) and Juhn and Mccue (2017), most of the production complementarities stressed by Becker (1991) have lost their central role in modern families. New household technologies (washing machines, vacuum cleaners, etc.) have reduced time devoted to household tasks, while the development of service industries has allowed individuals to buy in the market most of the goods (such as processed food) which used to be produced within the household. The only area where specialization has remained intact is with respect to children. Indeed, when considering individuals without children, there is no evidence that married women earn less than single women, while women with children have substantially lower earnings (Juhn and Mccue, 2017). SSM legalization should have therefore increased specialization within households only under the condition that it had an impact on fertility.

Overall, same-sex couples were less likely to have children: only 13% of male same-sex couples in the sample had a child living with them, compared to 28% of female same-sex couples and 58% of opposite-sex couples. Even when looking only at married couples, couples with children represented 22% of male same-sex couple, 36% of female same-sex couples, and 59% of opposite-sex couples.[24] Moreover, there seems to be more specialization among married couples than unmarried ones: when examining unmarried male same-sex couples, almost 70% of them have both partners working, compared with 63% among married male same-sex couples. Similar gaps can be found among female same-sex couples (67% vs. 63%) and opposite-sex couples (62% vs. 57%).[25]

As shown in Table 9, the effect of SSM legalization on the number of children in the household is close to zero and statistically insignificant both for gays (Column 1) and lesbians (Column 5). There is also no detectable effect on the probability of having a child (Columns 2 and 6). These conclusions do not change if lagged indicators of SSM legalization are used to allow couples more time to adjust their behaviors and fertility decisions (Columns 3 and 4 for gays, Columns 7 and 8 for lesbians). Restricting the sample size to consider only households in usual childbearing years also results in statistically insignificant coefficients.[26]

---

[24] See Section F.1 in the Online Appendix for additional descriptive statistics.
[25] See Section B in the Online Appendix.

To summarize, it is not surprising that SSM legalization did not lead to an increase in intra-household specialization since the main factor behind the advantages of home production – fertility – was not affected by this policy reform. It is true that married same-sex couples were more likely to have children and higher levels of intra-household specialization, but the policy reform itself does not seem to have trigger an increase in fertility among same-sex couples. Actually, the difference in fertility between married and unmarried same-sex couples predated the policy reform. The gap is mainly due to children older than five, thus implying than same-sex couple with a higher fertility rate were more likely to get married, not vice versa.

Finally, as discussed in Section F.2, couples with children were less likely to have both partners working, but SSM legalization positively affected more these couples than those without children, thus partially compensating the negative relation between fertility and employment. This result is in contrast with the hypothesis that marriage would provide the legal vehicle to further increase specialization within families with children.

**8.2 Additional potential mechanisms**

There are two additional reasons which may have explained a negative impact of SSM legalization. First, SSM may have led to an expansion of health insurance coverage among same-sex partners. Therefore, spouses who gained access to health insurance through their partners were no longer obliged to have a job in order to be covered by insurance. As discussed in Section F.2 in the Online Appendix, such access to health insurance depended not only on federal and state regulations, but also on case-by-case decisions by single employers and insurance companies. Therefore, given the resulting uncertainty, it is understandable that homosexual individuals did not overwhelmingly stop working just because they could potentially access their partners' health insurance.

Second, SSM legalization affected how same-sex married couples were taxed. Even if there are strong links between tax regulations and marital status (See Appendix F.3), these were not enforced until the U.S. Supreme Court decision in 2013. Moreover, married couples in the U.S. can decide whether to file taxes jointly or separately. Overall, given the "marriage penalty" in the U.S. tax system (Widiss, 2016), any tax incentive following the legalization of SSM might have attenuated the positive impact estimated in Tables 1 and 2.

There are instead three alternative reasons behind discrimination which could be reasonably behind the estimated positive effect of SSM. First, married individuals may have decided to start working more to save money and build a family later on. For instance, same-sex couples may have decided to save more in order to buy a house and/or have children in the following years. This hypothesis is not supported by the estimated null (or even negative) impact of SSM legalization on homeownership rates reported in Appendix F.4.

---

[26] See Section F.2 in the Online Appendix. In addition to this, Section F.2 also tests the existence of heterogeneities by income levels. The coefficient of SSM legalization remains statistically insignificant after including household income as controls. The interaction term between SSM legalization and income is positive, but its magnitude is not large enough to compensate the negative relation between fertility and household income.

Second, SSM legalization may have changed the composition and matching patterns among gays and lesbians, thus leading individuals with a higher propensity to work to match with similar partners. Nevertheless, there is no evidence that SSM legalization led to higher positive assortative matches among same-sex couples (Appendix F.5)

Third, a more tolerant environment may have improved mental health among LGBT members, thus improving productivity and indirectly leading to higher employment and earning levels. As discussed in Appendix F.6, several researchers have already documented the improvements in mental health following SSM legalization, but it is not possible to test with the available data whether this actually led to substantial higher productivity.

## 9. Limitations and future research

This study has two clear limitations due to data availability. First, homosexual individuals are identified only though their relation with the household head. Therefore, this paper has focused on cohabitating same-sex couples. There is no information available in the ACS on gays and lesbians without a partner, same-sex couples not living together, or bisexual individuals married with an opposite-sex individual. While the empirical analysis has tested the validity of the main results to potential compositional changes, it would be interesting by itself to investigate the impact of SSM legalization on these additional categories, as well as to other members of the LGBT community. Unfortunately, such an extension is not possible with the current data available in the U.S.

The second limitation is the lack of longitudinal data. While difference-in difference estimates can be computed using repeated cross-sections, several extensions are limited by the fact that individuals are not followed over time. For instance, it is not possible to investigate whether the wage increase documented for gay men in Table 3 is due to minority workers being more likely to get a raise or a promotion within the same job, or because these workers changed jobs, maybe switching to a less tolerant but more remunerative occupation. Future waves of the Panel Study of Income Dynamics (PSID) should identify same-sex couples, so hopefully a deeper evaluation will be possible in a few years.

Future research could also extend this analysis by looking at additional economic outcomes and how these variables have been impacted by same-sex marriage legalization. For instance, due to lack of data and small sample sizes, there is not enough information on how time use differs between same-sex and opposite-sex couples, and whether such time allocation has changed once same-sex couples have been allowed to get married. Furthermore, it would be interesting to understand if and how same-sex couples specialize and, once they have children, which partner spends more time taking care of them.

## 10. Conclusions

This paper exploits variations across U.S. states in the different timing of same-sex marriage reforms to show that this policy was associated with an improvement in labor market outcomes among same-sex couples. The second part of the analysis focuses on providing a theoretical framework and evidence supporting the hypothesis that SSM legalization reduced discrimination towards sexual minorities, thus boosting earnings and employment. This result is in line with experimental and observational studies showing that prejudiced attitudes towards LGBT individuals can be effectively and durably reduced (Broockman and Kalla, 2016; (Aksoy et al. 2018), thus suggesting that individual preferences and social norms might be changed through legal and economic reforms.

From a policy perspective, these results emphasize the overwhelming positive effects of extending civil and human rights to sexual minorities. This paper provides an economic rationale to marriage equality. Analogously to the increases in female labor force participation witnessed in the past decades, legalizing same-sex marriage led to higher integration of same-sex couples in the labor market, thus potentially stimulating economic growth and a more efficient allocation of human capital.

Even if accurate estimates of the LGBT population are difficult to obtain,[27] Stephens-Davidowitz (2017) combined different online data to argue that 5% of men are gay in the U.S., and a substantial share of them has not come out yet (or they are even married to a female partner). Recent surveys found that 7.5% of Millennials identify themselves as non-heterosexual (Gates, 2017). Therefore, LGBT individuals represent a sizable portion of the U.S. population: lower discrimination based on sexual orientation and higher employment among gays and lesbians are likely to have positive macroeconomic effects (and improve matching based on sexual preferences in the marriage market).

---

[27] See Section A.2.2 in the Online Appendix for an overview of different estimates.

Journal of Public Economics*, 118, 26–40.

---. (2017): *Everybody Lies: Big Data, New Data, and What the Internet Can Tell Us About Who We Really Are*, Dey Street Books.

STEVENSON, B. (2007): "The Impact of Divorce Laws on Marriage-Specific Capital," *Journal of Labor Economics*, 25, 75–94.

STEVENSON, B., AND J. WOLFERS. (2006): "Bargaining in the Shadow of the Law: Divorce Laws and Family Distress," *Quarterly Journal of Economics*, 121, 267–88.

---. (2007): "Marriage and Divorce : Changes and their Driving Forces," *The Journal of Economic Perspectives*, 21, 27–52.

TANKARD, M. E., AND E. L. PALUCK. (2017): "The Effect of a Supreme Court Decision Regarding Gay Marriage on Social Norms and Personal Attitudes," *Psychological Science*, 28, 1334–44.

TRANDAFIR, M. (2015): "Legal Recognition of Same-Sex Couples and Family Formation," *Demography*, 52, 113–51.

U.S. CENSUS. (2013): "Frequently Asked Questions About Same-Sex Couple Households," *U.S. Census*, August.

---. (2017): "Mandatory vs. Voluntary Methods," *ACS Methodology*, .

WEICHSELBAUMER, D. (2003): "Sexual orientation discrimination in hiring," *Labour Economics*, 10, 629–42.

WIDISS, D. A. (2016): "Legal recognition of same-sex relationships: New possibilities for research on the role of marriage law in household labor allocation," *Journal of Family Theory and Review*, 8, 10–29.

YOUNG, A. (2017): "Consistency without Inference: Instrumental Variables in Practical Application," *LSE working paper*, .
31

**Tables**

**Table 1: Effect of SSM legalization on probability both partners working.**

|  | Same-sex couples | | | | | | |
|---|---|---|---|---|---|---|---|
|  | 2008-16 | | | | | 2012-16 | 2014-16 |
|  | Male | | Female | | Male and female | | |
|  | (1) | (2) | (3) | (4) | (5) | (6) | (7) |
| SSM legal | 0.020* | 0.023** | 0.018* | 0.023** | 0.024*** | 0.044*** | 0.060*** |
|  | (0.012) | (0.011) | (0.011) | (0.010) | (0.008) | (0.014) | (0.009) |
| Year FE | Yes | Yes | Yes | Yes | Yes | Yes | Yes |
| State FE | Yes | Yes | Yes | Yes | Yes | Yes | Yes |
| Linear state trends | Yes | Yes | Yes | Yes | Yes | Yes | Yes |
| Quadratic state trends | Yes | Yes | Yes | Yes | Yes | Yes | Yes |
| Individual controls | Yes | Yes | Yes | Yes | Yes | Yes | Yes |
| State controls | No | Yes | No | Yes | Yes | Yes | Yes |
| Observations | 28,118 | 28,118 | 29,796 | 29,796 | 57,914 | 35,991 | 23,242 |
| Number clusters | 51 | 51 | 51 | 51 | 51 | 51 | 51 |
| Average dep var | 0.668 | 0.668 | 0.660 | 0.660 | 0.664 | 0.662 | 0.666 |
| Adjusted $R^2$ | 0.107 | 0.108 | 0.098 | 0.098 | 0.102 | 0.105 | 0.110 |

This table analyzes whether same-sex couples were more likely to have both partners working after the legalization of same-sex marriage. Difference-in-difference (DiD) estimates. Standard errors in parenthesis clustered at the state level. Individual and household controls: age, education, ethnicity, race and language of household head and partner. Household controls also include interaction terms between household head's and partner's age, as well as education. State controls: unemployment rate, income per capita, racial and age composition, percentage of state population with positive welfare (public assistance) income, and cohabitation rate among opposite-sex couples. Source: ACS 2008-2016 for Columns 1-5, ACS 2012-2016 for Column 6, ACS 2014-2016 for Column 7.
* $p < 0.10$, ** $p < 0.05$, *** $p < 0.01$

**Table 2: Effect of SSM legalization on employment.**

|  | Male and female same-sex couples | | | |
|---|---|---|---|---|
|  | (1) | (2) | (3) | (4) |
|  | HH hours | Both 40h | Both 30h | HH gap |
| SSM legal | 1.294*** | 0.030*** | 0.025*** | -0.936*** |
|  | (0.483) | (0.008) | (0.009) | (0.340) |
| Year FE | Yes | Yes | Yes | Yes |
| State FE | Yes | Yes | Yes | Yes |
| Linear state trends | Yes | Yes | Yes | Yes |
| Quadratic state trends | Yes | Yes | Yes | Yes |
| Individual controls | Yes | Yes | Yes | Yes |
| State controls | Yes | Yes | Yes | Yes |
| Observations | 57,815 | 57,914 | 57,914 | 57,815 |
| Number clusters | 51 | 51 | 51 | 51 |
| Average dep var | 69.40 | 0.461 | 0.612 | 15.40 |
| Adjusted $R^2$ | 0.151 | 0.078 | 0.097 | 0.020 |

This table analyzes the effect of legalizing same-sex marriage on the employment of same-sex couples. The dependent variables are: total number of hours worked weekly by the household head and the partner (Column 1), whether both the household head and the partner worked more than 40 hours per week (Column 2), whether both the household head and the partner worked more than 30 hours per week (Column 3), absolute value of the difference in hours worked weekly by the household head and the partner (Column 4). Male and female same-sex couples have been considered jointly. Difference-in-difference (DiD) estimates. Standard errors in parenthesis clustered at the state level. Individual and household controls: age, education, ethnicity, race and language of household head and partner. Household controls also include interaction terms between household head's and partner's age, as well as education. State controls: unemployment rate, income per capita, racial and age composition, percentage of state population with positive welfare (public assistance) income, and cohabitation rate among opposite-sex couples. Source: ACS 2008-2016.
* $p < 0.10$, ** $p < 0.05$, *** $p < 0.01$

**Table 3: Effect of SSM legalization on hourly earnings.**

|  | Same-sex couples | | |
|---|---|---|---|
|  | (1) | (2) | (3) |
|  | Male | Female | All |
| SSM legal | 0.028** | 0.006 | 0.014 |
|  | (0.014) | (0.010) | (0.010) |
| Year FE | Yes | Yes | Yes |
| State FE | Yes | Yes | Yes |
| Linear state trends | Yes | Yes | Yes |
| Quadratic state trends | Yes | Yes | Yes |
| Individual controls | Yes | Yes | Yes |
| State controls | Yes | Yes | Yes |
| Observations | 40,757 | 44,983 | 85,740 |
| Number of clusters | 51 | 51 | 51 |
| Average dep var | 3.16 | 3.00 | 3.08 |
| Adjusted $R^2$ | 0.267 | 0.302 | 0.286 |

This table analyzes the effect of legalizing same-sex marriage on the earnings (in logs) of same-sex couples. Self-employed individuals excluded when analyzing hourly earnings. Difference-in-difference (DiD estimates). Standard errors in parenthesis clustered at the state level. Individual and household controls: age, education, ethnicity, race and language of household head and partner. Individual controls also include whether the individual was identified as the household head or the partner. Household controls also include interaction terms between household head's and partner's age, as well as education. State controls: unemployment rate, income per capita, racial and age composition, percentage of state population with positive welfare (public assistance) income, and cohabitation rate among opposite-sex couples. Source: ACS 2008-2016.
* $p < 0.10$, ** $p < 0.05$, *** $p < 0.01$

**Table 4: Effect of SSM legalization on probability both partners working. Roommates.**

|  | Same-sex couples and roommates | | | |
|---|---|---|---|---|
|  | Male | Female | All | |
|  | (1) | (2) | (3) | (4) |
| SSM legal | 0.007 | 0.030*** | 0.018** | 0.024*** |
|  | (0.012) | (0.010) | (0.007) | (0.008) |
| Roommate | -0.030*** | -0.023** | -0.021*** |  |
|  | (0.006) | (0.009) | (0.006) |  |
| SSM legal * Roommate | 0.017 | 0.006 | 0.013 |  |
|  | (0.011) | (0.014) | (0.009) |  |
| Year FE | Yes | Yes | Yes | Yes |
| State FE | Yes | Yes | Yes | Yes |
| Linear state trends | Yes | Yes | Yes | Yes |
| Quadratic state trends | Yes | Yes | Yes | Yes |
| Individual controls | Yes | Yes | Yes | Yes |
| State controls | Yes | Yes | Yes | Yes |
| Observations | 41,389 | 38,135 | 79,524 | 79,524 |
| Number of clusters | 51 | 51 | 51 | 51 |
| Average dep var | 0.692 | 0.671 | 0.682 | 0.682 |
| Adjusted $R^2$ | 0.085 | 0.087 | 0.083 | 0.083 |

This table analyzes whether same-sex couples were more likely to have both partners working after the legalization of same-sex marriage. The sample includes same-sex married couples, unmarried couples, and roommates aged 30-60. Difference-in-difference (DiD estimates). Standard errors in parenthesis clustered at the state level. Individual and household controls: age, education, ethnicity, race and language of household head and partner/roommate. Household controls also include interaction terms between household head's and partner's (roommate's) age, as well as education. State controls: unemployment rate, income per capita, racial and age composition, percentage of state population with positive welfare (public assistance) income, and cohabitation rate among opposite-sex couples. Source: ACS 2008-2016.
* $p < 0.10$, ** $p < 0.05$, *** $p < 0.01$

**Table 5: Effect of same-sex laws on probability both partners working.**

|  | Male and female same-sex couples | | | | |
| --- | --- | --- | --- | --- | --- |
|  | (1) | (2) | (3) | (4) | (5) |
| SSM legal | 0.024*** | 0.024*** | 0.024*** | 0.025*** | 0.023*** |
|  | (0.008) | (0.008) | (0.008) | (0.008) | (0.008) |
| SSM ban |  | -0.068*** | -0.071*** | -0.064*** | -0.087*** |
|  |  | (0.015) | (0.015) | (0.015) | (0.018) |
| Domestic partnership |  |  | 0.069*** | 0.071*** | 0.073*** |
|  |  |  | (0.024) | (0.025) | (0.025) |
| Civil union |  |  | 0.007 | 0.011 | 0.012 |
|  |  |  | (0.017) | (0.018) | (0.018) |
| No discrimination |  |  |  | 0.186*** | 0.180*** |
|  |  |  |  | (0.057) | (0.059) |
| No discrimination public employees |  |  |  | 0.026 | 0.035 |
|  |  |  |  | (0.022) | (0.022) |
| Second-parent adoption |  |  |  |  | 0.270*** |
|  |  |  |  |  | (0.094) |
| No adoption by same-sex couples |  |  |  |  | -0.063 |
|  |  |  |  |  | (0.043) |
| Year FE | Yes | Yes | Yes | Yes | Yes |
| State FE | Yes | Yes | Yes | Yes | Yes |
| Linear state trends | Yes | Yes | Yes | Yes | Yes |
| Quadratic state trends | Yes | Yes | Yes | Yes | Yes |
| Individual controls | Yes | Yes | Yes | Yes | Yes |
| State controls | Yes | Yes | Yes | Yes | Yes |
| Observations | 57,914 | 57,914 | 57,914 | 57,914 | 57,914 |
| Number of clusters | 51 | 51 | 51 | 51 | 51 |
| Average dep var | 0.664 | 0.664 | 0.664 | 0.664 | 0.664 |
| Adjusted $R^2$ | 0.102 | 0.102 | 0.102 | 0.102 | 0.102 |

This table analyzes whether same-sex couples were more likely to have both partners working after the introduction of laws concerning same-sex individuals. Difference-in-difference (DiD estimates). A second-parent adoption is a legal procedure that allows a same-sex parent to adopt his or her partner's biological or adoptive child without terminating the legal rights of the first parent. Standard errors in parenthesis clustered at the state level. Individual and household controls: age, education, ethnicity, race and language of household head and partner. Household controls also include interaction terms between household head's and partner's age, as well as education. State controls: unemployment rate, income per capita, racial and age composition, percentage of state population with positive welfare (public assistance) income, and cohabitation rate among opposite-sex couples. Source: ACS 2008-2016.
* $p < 0.10$, ** $p < 0.05$, *** $p < 0.01$

**Table 6: Effect of SSM legalization on probability both partners working. Marital status.**

|  | Same-sex married and unmarried couples | | |
| --- | --- | --- | --- |
|  | Male | Female | All |
|  | (1) | (2) | (3) |
| SSM legal | 0.017 | 0.049** | 0.035** |
|  | (0.019) | (0.019) | (0.014) |
| Married | -0.054*** | -0.055*** | -0.054*** |
|  | (0.014) | (0.014) | (0.009) |
| SSM legal * Married | 0.028* | 0.024* | 0.024** |
|  | (0.016) | (0.014) | (0.009) |
| Year FE | Yes | Yes | Yes |
| State FE | Yes | Yes | Yes |
| Linear state trends | Yes | Yes | Yes |
| Quadratic state trends | Yes | Yes | Yes |
| Individual controls | Yes | Yes | Yes |
| State controls | Yes | Yes | Yes |
| Observations | 17,558 | 18,433 | 35,991 |
| Number of clusters | 51 | 51 | 51 |
| Average dep var | 0.671 | 0.654 | 0.662 |
| Adjusted $R^2$ | 0.114 | 0.100 | 0.107 |

This table analyzes whether same-sex couples were more likely to have both partners working after the legalization of same-sex marriage. Difference-in-difference (DiD estimates). Standard errors in parenthesis clustered at the state level. Individual and household controls: age, education, ethnicity, race and language of household head and partner. Household controls also include interaction terms between household head's and partner's age, as well as education. State controls: unemployment rate, income per capita, racial and age composition, percentage of state population with positive welfare (public assistance) income, and cohabitation rate among opposite-sex couples. Source: ACS 2012-2016.
* $p < 0.10$, ** $p < 0.05$, *** $p < 0.01$

**Table 7: Effect of SSM legalization on Google searches for *Leviticus*.**

|  | Web search intensity | | | | | |
|---|---|---|---|---|---|---|
|  | (1) | (2) | (3) | (4) | (5) | (6) |
| SSM legal | -2.094*** | -1.974** | -1.373* | -1.365* | -1.323* | -1.150 |
|  | (0.736) | (0.809) | (0.764) | (0.761) | (0.769) | (1.130) |
| SSM legal (Lag 1) |  |  | -1.130 | -1.143 | -0.636 | -0.438 |
|  |  |  | (0.822) | (0.834) | (0.809) | (0.967) |
| SSM legal (Lag 2) |  |  | -4.010*** | -3.974*** | -2.483** | -1.827 |
|  |  |  | (0.898) | (0.907) | (1.074) | (1.199) |
| Year FE | Yes | Yes | Yes | Yes | Yes | Yes |
| Linear state trends | No | No | No | No | Yes | Yes |
| Quadratic state trends | No | No | No | No | No | Yes |
| State controls | Yes | Yes | Yes | Yes | Yes | Yes |
| Policy controls | No | Yes | Yes | Yes | Yes | Yes |
| LGBT searches | No | No | No | Yes | Yes | Yes |
| Observations | 663 | 663 | 663 | 663 | 663 | 663 |
| Number of clusters | 51 | 51 | 51 | 51 | 51 | 51 |
| Average dep var | 20.87 | 20.87 | 20.87 | 20.87 | 20.87 | 20.87 |
| Within $R^2$ | 0.597 | 0.606 | 0.620 | 0.620 | 0.729 | 0.761 |
| Overall $R^2$ | 0.081 | 0.069 | 0.065 | 0.060 | 0.129 | 0.138 |

This table analyzes whether search intensities on Google for the word *Leviticus* changed after the legalization of same-sex marriage. Difference-in-difference (DiD estimates). State-level analysis with state fixed effects. Robust standard errors in parenthesis. State controls: unemployment rate, income per capita, racial and age composition, percentage of state population with positive welfare (public assistance) income, and cohabitation rate among opposite-sex couples. Policy controls: constitutional ban on same-sex marriage, legalization domestic partnership and civil union, anti-discrimination laws, legalization or prohibited adoptions by same-sex couples. LGBT searches measures the intensity of Google searches on LGBT topics. Source: ACS and Google Trends 2004-2016.
* $p < 0.10$, ** $p < 0.05$, *** $p < 0.01$

**Table 8: Effect of SSM legalization on occupation.**

|  | Male and female same-sex couples | | | | |
|---|---|---|---|---|---|
|  | Share women > 0.5 | | | Share women | Self-Empl |
|  | Head and Partner | Head | Employed | | |
|  | (1) | (2) | (3) | (4) | (5) |
| SSM legal | -0.014** | -0.019** | -0.012* | -0.006* | -0.019*** |
|  | (0.006) | (0.009) | (0.007) | (0.003) | (0.006) |
| Year FE | Yes | Yes | Yes | Yes | Yes |
| State FE | Yes | Yes | Yes | Yes | Yes |
| Linear state trends | Yes | Yes | Yes | Yes | Yes |
| Quadratic state trends | Yes | Yes | Yes | Yes | Yes |
| Individual controls | Yes | Yes | Yes | Yes | Yes |
| State controls | Yes | Yes | Yes | Yes | Yes |
| Observations | 106,230 | 54,124 | 92,135 | 106,230 | 56,633 |
| Number of clusters | 51 | 51 | 51 | 51 | 51 |
| Average dep var | 0.530 | 0.504 | 0.526 | 0.534 | 0.175 |
| Adjusted $R^2$ | 0.010 | 0.009 | 0.009 | 0.011 | 0.035 |

This table analyzes the effect of legalizing same-sex marriage on the occupational choices of same-sex couples. Difference-in-difference (DiD estimates). The dependent variables are: whether the individual worked in an occupation in which more than 50% of employees are women (Columns 1-3), the share of employees within an occupation who are women (Column 4), whether the household head or the partner were self-employed (Column 5). Standard errors in parenthesis clustered at the state level. Individual and household controls: age, education, ethnicity, race and language of household head and partner. Individual controls also include whether the individual was identified as the household head or the partner (Columns 1, 3, 4). Household controls also include interaction terms between household head's and partner's age, as well as education. State controls: unemployment rate, income per capita, racial and age composition, percentage of state population with positive welfare (public assistance) income, and cohabitation rate among opposite-sex couples. Individuals without any work experience in the 5 years preceding the interview or who had never worked have been excluded in Columns 1-4. Source: ACS 2008-2016 and Census 2000.
* $p < 0.10$, ** $p < 0.05$, *** $p < 0.01$

**Table 9: Effect of SSM legalization on fertility.**

|  | Male same-sex couples | | | | Female same-sex couples | | | |
|---|---|---|---|---|---|---|---|---|
|  | (1) | (2) | (3) | (4) | (5) | (6) | (7) | (8) |
|  | NChild | Child | NChild | Child | NChild | Child | NChild | Child |
| SSM legal | 0.002 | 0.002 | -0.005 | -0.002 | -0.011 | -0.000 | -0.003 | 0.003 |
|  | (0.019) | (0.009) | (0.020) | (0.009) | (0.019) | (0.010) | (0.021) | (0.012) |
| SSM legal (Lag 1) |  |  | -0.021 | -0.009 |  |  | 0.021 | 0.007 |
|  |  |  | (0.013) | (0.008) |  |  | (0.018) | (0.010) |
| SSM legal (Lag 2) |  |  | -0.012 | -0.008 |  |  | 0.017 | 0.007 |
|  |  |  | (0.015) | (0.008) |  |  | (0.020) | (0.012) |
| Year FE | Yes | Yes | Yes | Yes | Yes | Yes | Yes | Yes |
| State FE | Yes | Yes | Yes | Yes | Yes | Yes | Yes | Yes |
| Linear state trends | Yes | Yes | Yes | Yes | Yes | Yes | Yes | Yes |
| Quadratic state trends | Yes | Yes | Yes | Yes | Yes | Yes | Yes | Yes |
| Individual controls | Yes | Yes | Yes | Yes | Yes | Yes | Yes | Yes |
| State controls | Yes | Yes | Yes | Yes | Yes | Yes | Yes | Yes |
| Observations | 28,047 | 28,118 | 28,047 | 28,118 | 29,701 | 29,796 | 29,701 | 29,796 |
| Number of clusters | 51 | 51 | 51 | 51 | 51 | 51 | 51 | 51 |
| Average dep var | 0.215 | 0.125 | 0.215 | 0.125 | 0.456 | 0.281 | 0.456 | 0.281 |
| Adjusted $R^2$ | 0.053 | 0.055 | 0.053 | 0.055 | 0.080 | 0.087 | 0.080 | 0.087 |

This table analyzes the effect of legalizing same-sex marriage on the fertility of same-sex couples. The dependent variables are: number of children in the households (odd-numbered columns), and whether there is a child living in the household (even-numbered columns). Households with more than 4 children (top 1%) have not been considered in Columns 1 and 5. Male and female same-sex couples have been considered separately. Difference-in-difference (DiD estimates). Standard errors in parenthesis clustered at the state level. Individual and household controls: age, education, ethnicity, race and language of household head and partner. Household controls also include interaction terms between household head's and partner's age, as well as education. State controls: unemployment rate, income per capita, racial and age composition, percentage of state population with positive welfare (public assistance) income, and cohabitation rate among opposite-sex couples. Source: ACS 2008-2016.
* $p < 0.10$, ** $p < 0.05$, *** $p < 0.01$